\documentclass[epj,a4paper]{svjour}
\usepackage{multirow,rotating}
\usepackage[latin1]{inputenc}  
\usepackage{graphicx}

\graphicspath{{../Figures-EPS/}{.}{}}
\usepackage[]{color}
\usepackage[]{graphicx}
\usepackage[]{enumerate}
\usepackage[]{lineno}
\usepackage[square,comma,numbers,sort&compress]{natbib} 

\newcommand{\dresden}{1}
\newcommand{\infnmi}{2}
\newcommand{\infnge}{3}
\newcommand{\edinburgh}{4}
\newcommand{\infnpd}{5}
\newcommand{\lngs}{6}
\newcommand{\atomki}{7}
\newcommand{\infnto}{8}
\newcommand{\unimi}{9}
\newcommand{\roma}{10}
\newcommand{\infnna}{11}
\newcommand{\teramo}{12}

\begin{document}

\title{Neutron-induced background by an $\alpha$-beam incident on a deuterium gas target and its implications for the study of the $^2$H($\alpha$,$\gamma$)$^6$Li reaction at LUNA}
\author{
	M.\,Anders\inst{\dresden} 
	D.\,Trezzi \inst{\infnmi} \and
	A.\,Bellini \inst{\infnge} \and
	M.\,Aliotta \inst{\edinburgh} \and
	D.\,Bemmerer \inst{\dresden} \and
	C.\,Broggini \inst{\infnpd} \and
	A.\,Caciolli \inst{\infnpd} \and
	H.\,Costantini \inst{\infnge} \thanks{Present address: CPPM, Universit\'e d'Aix-Marseille, CNRS/IN2P3, Marseille, France} \and
	P.\,Corvisiero \inst{\infnge} \and
	T.\,Davinson \inst{\edinburgh} \and
	Z.\,Elekes \inst{\dresden} \and
	M.\,Erhard \inst{\infnpd} \thanks{Present address: PTB, Braunschweig, Germany} \and
	A.\,Formicola \inst{\lngs} \and
	Zs.\,F\"ul\"op \inst{\atomki} \and
	G.\,Gervino \inst{\infnto} \and
	A.\,Guglielmetti \inst{\unimi,\infnmi} \and
	C.\,Gustavino \inst{\roma} \thanks{Corresponding author, e-mail address carlo.gustavino@roma1.infn.it} \and
	Gy.\,Gy\"urky \inst{\atomki} \and
	M.\,Junker \inst{\lngs} \and
	A.\,Lemut \inst{\infnge} \thanks{Present address: LBNL, Berkeley, USA} \and
	M.\,Marta \inst{\dresden} \thanks{Present address: GSI, Darmstadt, Germany} \and
	C.\,Mazzocchi \inst{\infnmi} \thanks{Present address: University of Warsaw, Poland} \and
	R.\,Menegazzo \inst{\infnpd} \and
	P.\,Prati \inst{\infnge} \and
	C.\,Rossi Alvarez \inst{\infnpd} \and
	D.\,Scott \inst{\edinburgh} \and
	E.\,Somorjai \inst{\atomki} \and
	O.\,Straniero \inst{\infnna,\teramo} \and
	T.\,Sz\"ucs \inst{\atomki} 
	(LUNA collaboration)
	}                     
\institute{
	Helmholtz-Zentrum Dresden-Rossendorf (HZDR), Dresden, Germany 
	\and
	Istituto Nazionale di Fisica Nucleare (INFN), Sezione di Milano, Milano, Italy 
	\and
	Dipartimento di Fisica, Universit\`a di Genova, and INFN Sezione di Genova, Genova, Italy 
	\and
	SUPA, School of Physics and Astronomy, University of Edinburgh, Edinburgh, United Kingdom 
	\and
	INFN Sezione di Padova, Padova, Italy 
	\and 
	INFN, Laboratori Nazionali del Gran Sasso, Assergi, Italy 
	\and 
	Institute of Nuclear Research (ATOMKI), Debrecen, Hungary 
	\and
	Dipartimento di Fisica Sperimentale, Universit\`a di Torino, and INFN Sezione di Torino, Torino, Italy 
	\and 
	Universit\`a degli Studi di Milano, Milano, Italy 
	\and
	INFN Sezione di Roma "La Sapienza", Roma, Italy, 
	\and
	INFN Sezione di Napoli, Napoli, Italy 
	\and
	Osservatorio Astronomico di Collurania, Teramo, Italy 
}
\date{\today}

\abstract{
The production of the stable isotope $^6$Li in standard Big Bang nucleosynthesis has recently attracted much interest. Recent observations in metal-poor stars suggest that a cosmological $^6$Li plateau may exist. If true, this plateau would come in addition to the well-known Spite plateau of $^7$Li abundances and would point to a predominantly primordial origin of $^6$Li, contrary to the results of standard Big Bang nucleosynthesis calculations. Therefore, the nuclear physics underlying Big Bang $^6$Li production must be revisited. The main production channel for $^6$Li in the Big Bang is the $^2$H($\alpha$,$\gamma$)$^6$Li reaction. The present work reports on neutron-induced effects in a high-purity germanium detector that were encountered in a new study of this reaction. In the experiment, an $\alpha$-beam from the underground accelerator LUNA in Gran Sasso, Italy, and a windowless deuterium gas target are used. A low neutron flux is induced by energetic deuterons from elastic scattering and, subsequently, the $^2$H(d,n)$^3$He reaction. Due to the ultra-low laboratory neutron background at LUNA, the effect of this weak flux of 2-3\,MeV neutrons on well-shielded high-purity germanium detectors has been studied in detail. Data have been taken at 280 and 400\,keV $\alpha$-beam energy and for comparison also using an americium-beryllium neutron source. 
\PACS{
	{26.35.+c}{Big Bang nucleosynthesis} \and
	{25.55.-e}{$^3$H-, $^3$He-, and $^4$He-induced reactions} \and	
	{25.40.Fq}{Inelastic neutron scattering} \and	
	{29.30.Kv}{X- and gamma-ray spectroscopy} 
    }
}
\authorrunning{M. Anders {\it et al.} (LUNA collab.)}
\titlerunning{Neutron-induced background by an $\alpha$-beam incident on a deuterium gas target...}

\maketitle

\section{Introduction}

\subsection{Cosmology and $^6$Li}

For 30 years now, the cosmological lithium puzzle has frustrated the efforts of observers and cosmologists \cite[a recent review]{Fields11-ARNPS}. Observations in metal-poor stars consistently show a factor of two or three lower abundance of the main stable lithium isotope $^7$Li (isotopic abundance on Earth 92.41\%) than what is predicted by standard Big Bang nucleosynthesis, even though some possible stellar solutions have been suggested \cite{Korn06-Nature}. 

Recent observations of the second stable lithium isotope $^6$Li (isotopic abundance on Earth 7.59\%) in metal-poor stars have introduced a possible additional puzzle, this time concerning cosmological $^6$Li. The observed \linebreak $^6$Li/$^7$Li abundance ratio of about 0.05 \cite{Asplund06-ApJ} largely exceeds the standard Big Bang nucleosynthesis prediction. Even though many of the claimed $^6$Li detections may be in error \cite{Garcia09-AA,Steffen10-IAU}, for a few metal-poor stars \cite{Smith93-ApJ} there still seems to be a $^6$Li isotopic abundance of a few percent  \cite{Steffen12-LiC}. These observations are much higher than the predicted $^6$Li yield from standard Big Bang nucleosynthesis \cite{Serpico04-JCAP}. 

$^6$Li is much more easily depleted through nuclear reactions than its more abundant sister isotope $^7$Li \cite{Prantzos06-AA}, and non-cosmological scenarios for significant $^6$Li production all co-produce $^7$Li and thus worsen the $^7$Li problem \cite[a very recent example]{Iocco12-PRL}. Therefore, the detections of $^6$Li in very old stars raise the question of a possible cosmological production. However, with standard Big Bang nucleosynthesis producing much too little $^6$Li \cite{Serpico04-JCAP}, all cosmological scenarios yielding enough  $^6$Li involve non-standard physics \cite{Kusakabe06-PRD,Pospelov07-PRL,Jedamzik09-NJP,Pospelov10-ARNPS}. 
Before studying such exotic scenarios, it is important to first determine precisely how much $^6$Li can be produced in standard Big Bang nucleosynthesis. With this experimental baseline, any missing additional $^6$Li provided by non-standard approaches can be quantified.

\subsection{The nuclear physics of $^6$Li production}

The $^2$H($\alpha$,$\gamma$)$^6$Li reaction is the dominant nuclear reaction for $^6$Li production in standard Big Bang nucleosynthesis \cite{Serpico04-JCAP}. At the energies\footnote{Throughout this text, $E$ always denotes the center-of-mass energy, and $E_\alpha$ the  $\alpha$-beam energy in the laboratory.} relevant to Big Bang nucleosynthesis, $E$ $\approx$ 50-300\,keV, the $^2$H($\alpha$,$\gamma$)$^6$Li cross section is very small. Therefore, it has never been measured experimentally at such low energies, and theoretical predictions remain uncertain \cite{Marcucci06-NPA,Hammache10-PRC,Mukh11-PRC_6Li}. In the low-energy domain, the cross section $\sigma_{24}(E)$ is usually parameterized as the astrophysical S-factor $S_{24}(E)$ defined by
\begin{equation} \label{eq:Sfactordefinition}
S_{24}(E) = E \, \sigma_{24}(E) \, \exp(2\pi\eta(E))
\end{equation}
with $E$ the center-of-mass energy, and $2\pi\eta(E)$ the Sommerfeld parameter parameterizing the exponential-like energy dependence of the probability of tunneling through the Coulomb barrier  \cite{Rolfs88-Book}. 

The reaction has been studied previously by in-beam $\gamma$-spectrometry around the $E$ = 0.711\,MeV resonance \cite{Mohr94-PRC}. At even higher energies, there are data by in-beam detection of the $^6$Li reaction products \cite{Robertson81-PRL}. In both cases \cite{Mohr94-PRC,Robertson81-PRL}, an $\alpha$-beam had been incident on a deuterium gas target. A third experiment at very low energy, using a deuterated polyethylene target, just resulted in an upper limit \cite{Cecil96-PRC}.

Two attempts to determine the $^2$H($\alpha$,$\gamma$)$^6$Li cross section have been made using the Coulomb dissociation technique \cite{Kiener91-PRC,Hammache10-PRC}, in which an energetic $^6$Li beam is shot on a target of high nuclear charge. The time-reversed reaction $^6$Li($\gamma$,$\alpha$)$^2$H is then studied using virtual photons. The cross section of interest is obtained by applying the detailed balance theorem \cite{Blatt91-Book}. This method is especially sensitive to quadrupole (E2) transitions in the excited $^6$Li nucleus, largely neglecting dipole transitions. It is furthermore limited by possible background from non-Coulomb, i.e. nuclear breakup \cite{Baur96-ARNPS}. 

Such experiments have been carried out recently, using a 150\,MeV/$A$ $^6$Li beam and the KaoS spectrometer at GSI \cite{Hammache10-PRC}. It was found that nuclear breakup dominated the signal, so only an experimental upper limit could be derived \cite{Hammache10-PRC}, but the observed angular distributions seemed to support a theoretical excitation function developed in the same paper. It is stated that the previous Coulomb breakup data \cite{Kiener91-PRC}, which had been obtained at a much lower $^6$Li projectile energy of 25\,MeV/$A$, are probably even more strongly affected by nuclear breakup, so that also these data should be interpreted as upper limits \cite{Hammache10-PRC}. 

Therefore, direct data on the $^2$H($\alpha$,$\gamma$)$^6$Li cross section at Big Bang energies are still needed. The present work lays the foundation for such an experiment at the Laboratory for Underground Nuclear Astrophysics (LUNA, \cite{Costantini09-RPP,Broggini10-ARNPS}) 400\,kV accelerator.  

\subsection{Processes expected to take place in the target}

In order to prepare the planned $^2$H($\alpha$,$\gamma$)$^6$Li experiment, it is necessary to study in details the dominating background. Inside the target, mainly the following nuclear reactions are expected to take place:
\begin{eqnarray*}
\rm ^2H(\alpha,\gamma)^6Li & & Q = 1.474\,{\rm MeV} \; ({\rm R}1)  \\
\rm ^2H(\alpha,\alpha)^2H & \longrightarrow  \; \rm ^2H(d,n)^3He \; & Q = 3.269\,{\rm MeV} \;({\rm R}2) \\
\rm   & \longrightarrow \; \rm ^2H(d,p)^3H\: &  Q = 4.033\,{\rm MeV} \; ({\rm R}3)
\end{eqnarray*}
Along with the reactions, also their $Q$-values are listed. Reaction\,R1 is the main reaction to be studied. Reactions~R2 and R3 are parasitic reactions induced by elastically scattered deuterons. As these deuterons have a maximum energy of only 356\,keV (for $E_\alpha$ = 400\,keV), other deuteron-induced reactions are negligible due to suppression by the Coulomb barrier. However, there is no Coulomb barrier for the energetic neutrons released by reaction\,R2, and the protons released by reaction\,R3 have an average energy of 3\,MeV. 

Therefore, these protons and neutrons may give rise to further reactions on the structural material of the gas target system and on the detector material. The cross sections of these two conjugate reactions are similar, and they have nearly the same energy dependence at the energies relevant here \cite{Leonard06-PRC}. As a consequence, the easily detectable protons from reaction~R3 may in principle be used to monitor, in a relative manner, also the yield of reaction~R2. 

\subsection{Neutron-induced effects in germanium detectors}

There is a considerable body of literature on this problem. Soon after the advent of lithium-drifted germanium, or Ge(Li), detectors, the effects of neutrons on these devices has been studied  \cite{Chasman65-NIM,Bunting74-NIM}. The energy loss of recoiling germanium atoms in the germanium explains the characteristic triangular shape due to (n,n'$\gamma$) reactions in germanium \cite{Fehrenbacher96-NIMA,Fehrenbacher97-NIMA}. These effects have already been included in Monte Carlo simulations \cite{Ljungvall05-NIMA}, and the possible discrimination against unwanted neutron effects in a highly segmented detector array has been studied \cite{Atac09-NIMA}. The response of HPGe detectors to strong Am-Be neutron sources \cite{Abt08-EPJA,Mei08-PRC} and the effects of  cosmic-ray induced neutrons on low-background germanium detector systems \cite{Heusser93-NIMB,Heusser95-ARNPS,Wordel96-NIMA} have both been studied in detail before. 

In the present work, a detailed study of the effects of low neutron fluxes on an underground, well-shielded germanium detector is presented, using parasitic neutrons produced by the $^2$H(d,n)$^3$He reaction in the in-beam experiment and, subsequently, an americium-beryllium (Am-Be) neutron source. 

This work is organized as follows. The experimental setup is described in sec.\,\ref{sec:Setup}, the Monte Carlo simulations and their validation using an americium-beryllium neutron source in sec.\,\ref{sec:MonteCarlo}. The main observations with the silicon and the germanium detectors, detecting charged particles and $\gamma$ rays, respectively, during the in-beam experiments are summarized in secs.\,\ref{sec:SiliconData} and \ref{sec:GermaniumData}. The results and the effects for the $^2$H($\alpha$,$\gamma$)$^6$Li experiment at LUNA are discussed in sec.\,\ref{sec:Discussion}, and a summary is given in sec.\,\ref{sec:Summary}.

\begin{figure*}[]
\centering{\includegraphics[angle=0,width=0.9\textwidth,trim=0 0cm 0 0,clip]
	{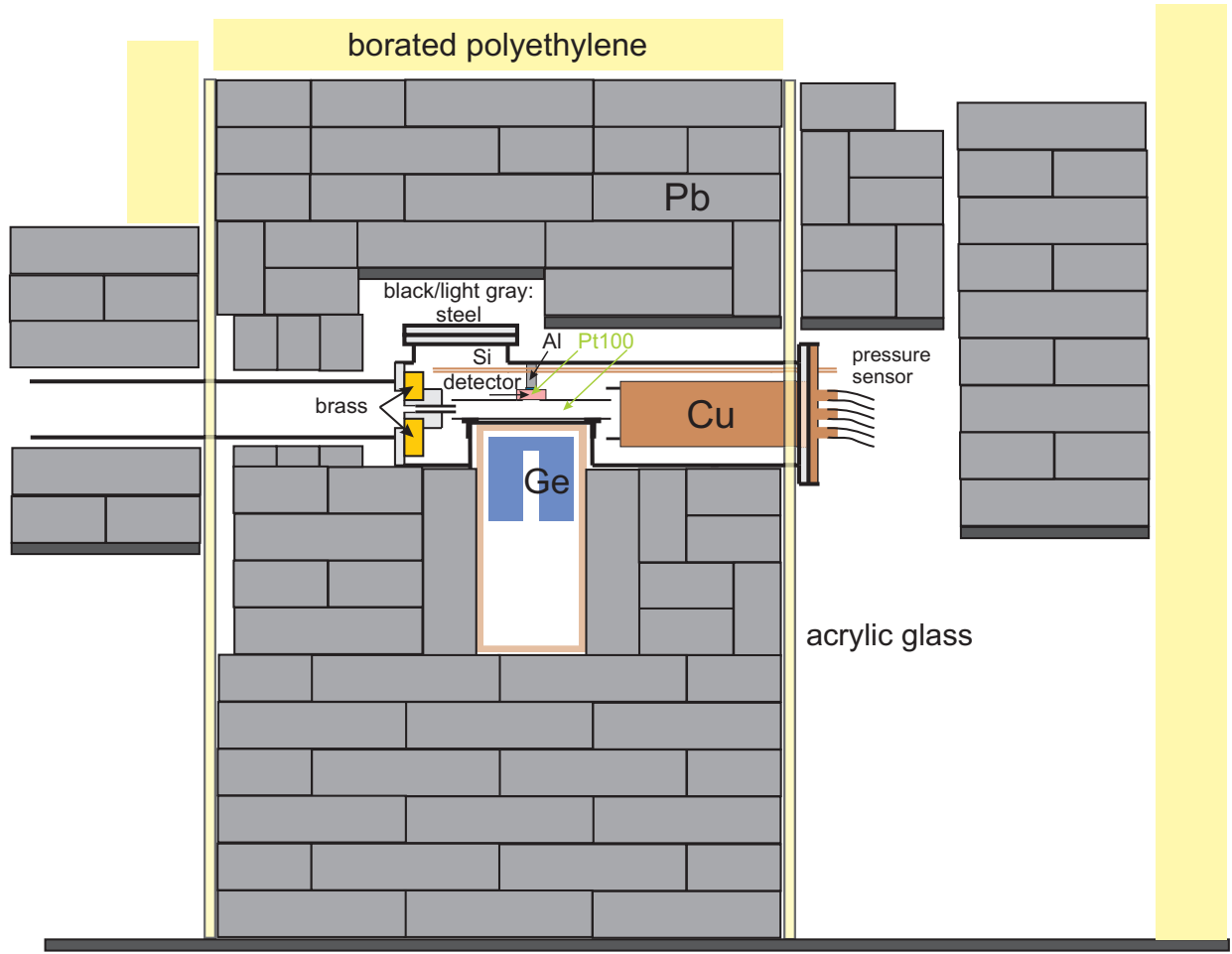}}
\caption{\label{fig:Setup} Experimental setup, as seen from the side. The central chamber of the windowless gas target is seen near the center of the plot. The $\alpha$-beam enters the target from the left through a 4\,cm long collimator of 7\,mm inner diameter and is stopped on a massive copper beam calorimeter. The germanium (for $\gamma$ rays) and silicon (for charged particles) detectors are also shown, as well as the Pt100 temperature sensors and the tube leading to the capacitive pressure sensor. The setup is surrounded by a lead castle and walls of borated polyethylene. The inner lead castle is surrounded by an anti-radon box made of acrylic glass.}
\end{figure*}

\section{Experimental setup}
\label{sec:Setup}

\subsection{The LUNA 400\,kV underground accelerator }

The experiment was performed at the Laboratory for Underground Nuclear Astrophysics (LUNA) 400\,kV  accelerator in the Gran Sasso laboratory, Italy (LNGS). The Gran Sasso underground facility is shielded from cosmic-ray induced muons by the Gran Sasso mountain, reducing the muon flux by six orders of magnitude \cite{MACRO90-PLB}. The ambient flux of energetic neutrons at Gran Sasso is of the order of 10$^{-6}$\,s$^{-1}$\,cm$^{-2}$ \cite{Belli89-NCA,Arneodo99-NCA}, three orders of magnitude lower than at the surface of the Earth.

LUNA is the world's only underground accelerator facility. It is dedicated to cross section measurements of astrophysically relevant reactions well below the Coulomb barrier \cite{Costantini09-RPP,Broggini10-ARNPS}, taking advantage of its ultra-low background \cite{Bemmerer05-EPJA,Caciolli09-EPJA,Szucs10-EPJA}. 

During the experiment described here, the LUNA \linebreak 400\,kV accelerator \cite{Formicola03-NIMA} provided a $^4$He$^+$ beam of 280 and 400\,keV energy, with an intensity of 100-350\,$\mu$A (typical value 300\,$\mu$A). 

\subsection{Windowless gas target system}

After the analyzing magnet, the $^4$He$^+$ beam was directed to a windowless gas target system consisting of three differential pumping stages \cite{Casella02-NIMA}. After passing through a series of three collimators of 25, 15, and 7\,mm diameter, the beam enters the target chamber (fig.\,\ref{fig:Setup}). The final collimator with 7\,mm diameter is directly at the entrance of the target chamber. The other two collimators are upstream at 69 and 51\,cm distance, respectively. Inside the target, deuterium gas of 99.8\% chemical and isotopic purity  was maintained at a constant pressure of 0.3\,mbar via a capacitive pressure sensor (independent of gas type, calibrated to 0.25\% precision) controlling the gas inlet via a feedback loop. 

The target chamber is a box-shaped container of 44\,cm length and 11$\times$13\,cm area, made of 2\,mm thick AISI 304 steel. 6\,cm downstream of the initial wall, a recess of 13\,cm length and 4.4\,cm height is introduced in order to insert the HPGe detector close to the beam line. At the center of the target chamber, a smaller, box-shaped AISI 304 steel inner tube of 1.8$\times$1.8\,cm area and 17\,cm length with a wall thickness of 1\,mm was introduced. This inner tube limits the lateral length that elastically scattered deuterons travel inside the gas, and therefore the deuterium target for the $^2$H(d,n)$^3$He reaction. The inner tube and the inner part of the collimator at the entrance of the target chamber were made from interchangeable parts, so they were replaced with a fresh component from time to time, typically after about $\sim$500\,C of beam charge. Due to proper inner linings, the position of the components was reproducible to better than 1\,mm.

Outside the target, the differential pumping system maintains already a medium vacuum (10$^{-3}$\,mbar) at the first pumping stage, evacuated by a 2050\,m$^3$/h Roots pump. The second pumping stage is equipped with three 1500\,l/s turbomolecular pumps, and the third pumping stage with a 360\,l/s turbomolecular pump. The vacuum in the second and third pumping stages is in the 10$^{-6}$-10$^{-7}$\,mbar range. The backing pumps for the three pumping stages are oil-free rotary vane compressors.

After traveling for 17.7\,cm through the gas target, the ion beam is stopped on the copper head of a beam calorimeter with constant temperature gradient to measure the beam intensity \cite{Casella02-NIMA}. Resistive heaters controlled via a feedback loop and forced cooling maintain a constant temperature gradient between 70\,$^\circ$C on the hot side, facing the beam, and 0\,$^\circ$C on the cold side of the calorimeter. 

The gas temperature inside the target chamber varied monotonously between 16\,$^\circ$C next to the water-cooled entrance collimator and 70\,$^\circ$C immediately adjacent to the hot side of the calorimeter. The pressure profile had been measured in a similar setup in a previous experiment at LUNA and was found to be flat \cite{Gyurky07-PRC}. It is assumed here that it is again flat. 

The effective target density depends on target pressure and temperature, but also on the beam heating effect \cite{Goerres80-NIM}. This effect has been studied previously at LUNA for $^3$He gas, in a setup similar to the present one, by double elastic scattering \cite{Marta06-NIMA}. For $^3$He gas, the elastic scattering data \cite{Marta06-NIMA} are in agreement with the predicted beam heating correction $\Delta T$ from the beam current $I$, energy loss in the target $dE/dx$, heat conductivity $\lambda$ and the approximate radii of target chamber $r_{\rm Chamber}$ and beam $r_{\rm Beam}$ using the following formula \cite{Osborne84-NPA}:
\begin{equation} \label{eq:Beamheating}
\Delta T = I\frac{dE}{dx} \frac{1}{2\pi \lambda} \ln \frac{r_{\rm Chamber}}{r_{\rm Beam}} 
\end{equation}
Using $I$ = 200-300\,$\mu$A, $dE/dx$ = 0.17 (0.19)\,keV/cm at 400 (280) keV, $\lambda$ = 1.31 mW/(K\,cm),
$r_{\rm Chamber}$ $\approx$ 9\,mm, $r_{\rm Beam}$ $\approx$ 3.5\,mm, a correction of 4-7\,K is obtained, corresponding to a 1-2\% reduction in effective target density. As in previous work \cite{Marta06-NIMA}, a conservative relative uncertainty of 40\% is assumed for this correction.

\subsection{Silicon particle detector}
\label{subsec:Silicon}

For the detection of protons from reaction~R3, a 1500\,$\mu$m thick silicon detector of 450\,mm$^2$ active area is installed atop the center of the gas target chamber (fig.\,\ref{fig:Setup}). At this position, it can detect protons from reaction R3 taking place both within the gas volume and in the wall of the inner tube, within the limits of the detector solid angle.
The silicon detector is covered with a 25\,$\mu m$ thick aluminum foil in order to stop elastically scattered $^4$He and $^2$H particles (with a range 1.5 and 3\,$\mu$m, respectively) while letting the energetic protons pass. Without the foil, the measured energy resolution of the detector itself is about 25\,keV full width at half maximum (FWHM) for 5.5\,MeV $\alpha$ particles from an $^{241}$Am source.

Without  active cooling and at high beam intensities, the silicon detector reaches a temperature of up to 45$\,^{\circ}\mathrm{C}$, as measured by a Pt100 resistor at the detector enclosure. This is so, because convective cooling is negligible at the present low target gas pressure. This increase in temperature leads to degraded performance of the detector, because the main contributor to its energy resolution is thermal noise.

Therefore, an active cooling is provided by a Peltier element on the back side of the detector. Its hot side is thermally connected by an aluminum bar to the target chamber wall (fig.\,\ref{fig:Setup}). When cooled by the Peltier element, the silicon detector temperature never exceeds 23$\,^{\circ}$$\mathrm{C}$, the recommended working temperature specified in the detector data sheet. 

\subsection{Germanium $\gamma$-ray detector and lead shield}
\label{subsec:Germanium}

The $\gamma$ rays are detected in a large high-purity germanium detector of 137\% relative efficiency. This detector is placed inside the recess in the target chamber (fig.\,\ref{fig:Setup}). The symmetry axis of the HPGe detector is 7.5\,cm downstream of the entrance of the gas target, and its end cap just 1.5\,cm below the beam axis. This detector has been optimized for ultra-low background, and it has previously been used to construct a setup with the lowest background ever observed for an in-beam $\gamma$-spectrometry setup \cite{Caciolli09-EPJA}. 

\begin{figure}[tb]
\includegraphics[angle=0,width=\columnwidth]{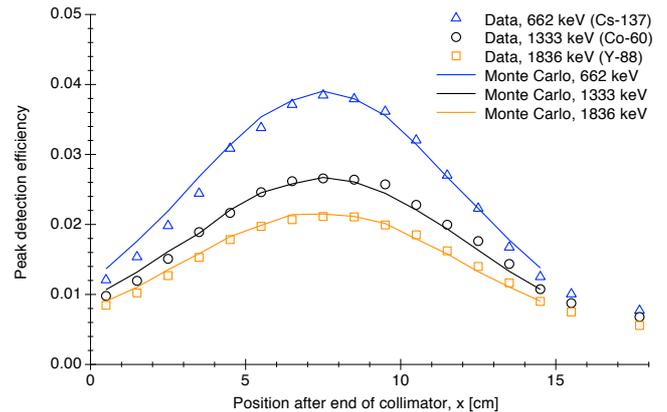}
\caption{\label{fig:Gamma-Efficiency} Full-energy peak detection efficiency measured with several calibrated $\gamma$-ray sources, as a function of the position $x$ after the end of the final collimator. The efficiency from the GEANT4 Monte Carlo simulation is also shown for several positions.
}
\end{figure}

The $\gamma$-ray detection efficiency was measured with calibrated $^{137}$Cs, $^{60}$Co, and $^{88}$Y sources (fig.\ref{fig:Gamma-Efficiency}) as a function of their position along the beam axis. Due to the close geometry used, the data had to be corrected for the true-co\-in\-cidence summing out effect, a correction amounting to up to 12\% at the closest distance studied here.

The typical resolution observed for the 1.333\,MeV line of $^{60}$Co is 2.7\,keV when the gas target pumps were switched off, a few tenths of a keV worse when the pumps were running. For the same $\gamma$-line, with the source placed near the center of the detector ($x$ = 7.5\,cm) a peak to Compton ratio of 65 was found. 

The detector was shielded from environmental radionuclides by a lead shield of at least 20\,cm thickness (fig.\,\ref{fig:Setup}). The lead is selected for a $^{210}$Pb content of just 25\,Bq/kg \cite{Caciolli09-EPJA}. In order to limit the impact of the neutrons, the setup was enclosed by a 5\,cm thick shield of borated polyethylene (fig.\,\ref{fig:Setup}), in order to exclude any possible background for other experiments in the Gran Sasso underground facility. This shield attenuated the additional neutron background produced by the present experiment outside the LUNA experimental hall to a level that was well below the already ultra-low ambient neutron flux at Gran Sasso \cite{Belli89-NCA,Arneodo99-NCA}. In order to reduce the background from radon decay products, the setup was enclosed in an anti-radon box made of acrylic glass that is flushed with radon-free nitrogen evaporated from a Dewar vessel. 

\subsection{Data acquisition}

The data from the germanium detector were passed on to two independent data acquisition systems for processing and storage. One branch of the data included the TNT2 digital data acquisition system in a CAEN N1728B module, operated in list mode. The dead time was estimated using a pulser. 

In the second branch, histograms were recorded at regular intervals using the Ortec 919E EtherNIM analog to digital converter and multichannel buffer. Here, the Gedcke-Hale algorithm was used for dead time estimation. The data from the silicon detector were treated just by the Ortec branch, with a dead time of less than 1\% estimated with a pulser connected to the preamplifier test input. 

The main parameters of the gas target system and calorimeter, such as the pressures observed in the various pumping stages and the temperatures at several places inside the calorimeter, were logged at intervals of a few seconds via a Labview-based slow control system.

\section{Monte Carlo simulation}
\label{sec:MonteCarlo}

The experimental setup was coded in two versions of \linebreak GEANT, each of them suited to the special needs of the case: A specially adapted version of GEANT3 taking into account the correct energy loss for charged particles was used to model the data in the silicon particle detectors. The details of this code have been discussed previously \cite{Arpesella95-NIMA}, and the output from the GEANT3 simulation is compared with the data in sec.\,\ref{sec:SiliconData}.

The response to $\gamma$ rays and neutrons was coded in GEANT4 \cite{Agostinelli03-NIMA}. The GEANT4 simulation is described in the present section.

\subsection{GEANT4 simulation}

The GEANT4 version used was 4.9.2, and the \linebreak {\tt QGSP\_BIC\_HP} physics list was activated. This physics list uses the GEANT4 binary cascade ({\tt BIC}) mechanism for primary protons and neutrons with energies below 10\,GeV. The excited nucleus created by the primary nucleon is then passed to the precompound model, which describes the nuclear de-excitation. For neutrons below 20\,MeV down to thermal energies, the data-driven high-precision neutron package ({\tt NeutronHP}) is used. The {\tt NeutronHP} version used was {\tt G4NDL3.13}, which is based on the ENDF/B-VI and JENDL libraries. The high-energy interactions are taken from the JENDL High Energy File 2007. 

The first step of the GEANT4 simulation was to adjust the details of the geometry for the HPGe detector used here, starting from the nominal geometry, until the data from the radioactive $\gamma$-ray sources $^{137}$Cs, $^{60}$Co, and $^{88}$Y placed at various positions along the beam path were reproduced to a precision of better than 10\%. The crystal diameter and length were reduced after a repair and etching process of the detector used here. In order to reproduce the measured efficiency profile (fig.\ref{fig:Gamma-Efficiency}), the following values were applied, with the data sheet values for the detector before repair given in brackets: Crystal diameter 86\,mm (91\,mm), length 85\,mm (91\,mm), distance to the casing 12\,mm (4\,mm). The quoted relative efficiency of the detector before (after) repair was 150\% (137\%). With the adopted dimensions given here, the simulation predicts a relative efficiency of 130\%, close to the data sheet efficiency after repair. 

For a $^{60}$Co source, a peak to Compton ratio of 64 was found in the simulation near the center of the target ($x$ = 7.5\,cm, with $x$ the position measured after the end of the final collimator), after scaling to obtain the same resolution as in the experiment. This is consistent with the experimental value of 65 (sec.\,\ref{subsec:Germanium}), showing that the simulated geometry seems to be appropriate. Also at other positions and for other sources, the efficiency data with $\gamma$-ray sources was well reproduced by the Monte Carlo simulation (fig.\ref{fig:Gamma-Efficiency}).  

As a second step, the neutron-induced background in the in-beam experiment was modeled in the given geometry. This was done in several discrete steps, in order to limit the computation time:

\begin{enumerate}
\item $^4$He$^+$ ions of the correct laboratory energy (280 or 400 keV) are shot into the gas target, and the Rutherford scattering reaction $^2$H($\alpha$,$\alpha$)$^2$H is implemented. The energy spectrum of emitted deuterons, integrated over all angles, is recorded and kept as input for the following step.
\item A source of energetic deuterons, distributed in energy according to their previously derived energy spectrum, is created. The point of emission is a cylinder of the dimensions of the He$^+$ beam, along the beam axis between final collimator and beam stop. The radial profile of emission points of origin is randomly distributed with a Gaussian distribution following the beam profile, and the angle of emission is also randomly distributed. The deuterium gas target is then irradiated with this source, and the neutron-emitting reaction $^2$H(d,n)$^3$He is  implemented. The interaction of the resulting neutrons with the structural and shielding materials and with the germanium detector material is then simulated.
\item The energy deposited in the HPGe detector by incident $\gamma$ rays is then recorded and histogrammed. The energy deposited by recoiling germanium atoms inside the detector crystal is also included, taking into account quenching effects \cite{Ljungvall05-NIMA}.
\end{enumerate}

\subsection{Benchmarking of the GEANT4 simulation using an americium-beryllium neutron source}
\label{subsec:AmBe}

In order to benchmark the GEANT4 simulation, a weak americium-beryllium (Am-Be) neutron source, emitting MeV neutrons by the $^9$Be($\alpha$,n)$^{12}$C reaction, was introduced to the setup. The $^{241}$Am activity was 185\,kBq, leading to an estimated neutron source strength of 13\,n/s. The $^{241}$Am nitrate was contained in a thin layer between two 0.1\,mm thick beryllium disks of 11\,mm diameter, housed in a 5\,mm thick polyethylene cylinder of 18\,mm diameter. 

\begin{figure}[tb]
\includegraphics[angle=-90,width=\columnwidth]{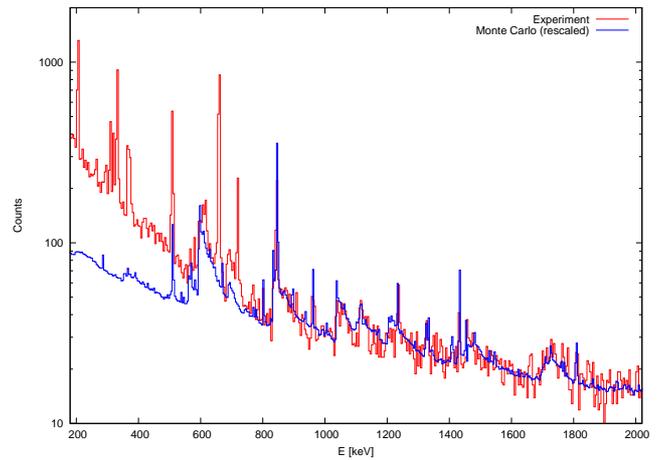} 
\caption{\label{fig:AmBe} Spectrum of the AmBe neutron source from the experiment and from the Monte Carlo simulation.}
\end{figure}

The Am-Be source was placed in the center of the target chamber. 
Due to restrictions on the use of neutron sources in the LNGS underground facility, the running time was only 10 hours, limiting the statistical precision of the data. Therefore, the data had to be rebinned in order to make the spectra comparable, and any comparison by necessity concentrates on the general features of the spectrum (fig.\,\ref{fig:AmBe}).

The triangular feature at 691\,keV is not well reproduced in the simulation, for reasons discussed below (sec.\,\ref{sec:Discussion}).  
At lower energies, $E_\gamma$ $<$ 500\,keV, the simulation significantly underpredicts the data. This is due to the Compton continuum from the $\gamma$ rays at 662 and 722\,keV that are emitted by the $^{241}$Am in the Am-Be source. These weak $\gamma$-ray branches of $^{241}$Am are not included in the Monte Carlo simulation. 
In addition to energetic neutrons from the $^9$Be($\alpha$,n)$^{12}$C reaction, an Am-Be source also emits Doppler-broadened 4.4\,MeV $\gamma$ rays from the decay of the first excited state of $^{12}$C. These $\gamma$ rays were not included in the Monte Carlo simulation, but they have only a very limited influence near the ROIs for the $^2$H($\alpha$,$\gamma$)$^6$Li reaction. 

Overall, the spectrum from the Am-Be source is well reproduced by the simulation, including the neutron triangles except for the one at 691\,keV where there seem to be problems with the treatment of the internal conversion process in GEANT4 (sec.\,\ref{sec:Discussion}).

\section{In-beam observations with the silicon particle detector}
\label{sec:SiliconData}

\begin{figure}[tb]
\includegraphics[angle=0,width=\columnwidth,trim=1.8cm 0 1.8cm 0,clip]{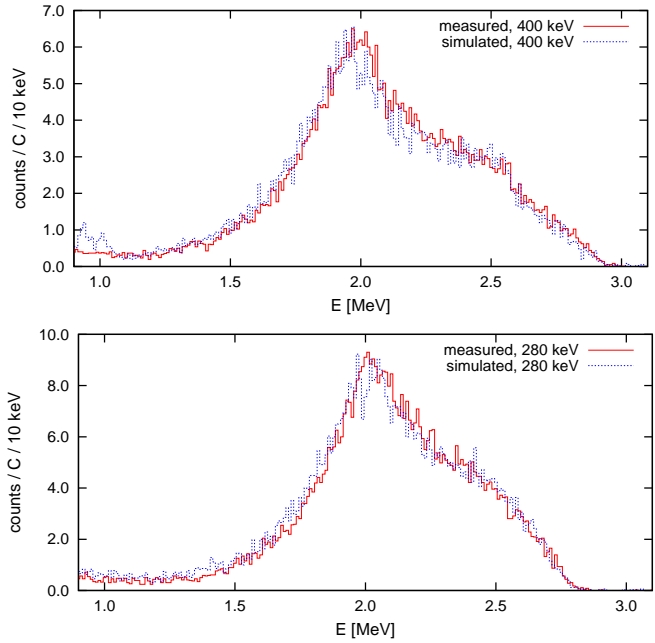}
\caption{\label{fig:Silicon} Experimental (red solid line) and GEANT3-simulated (blue dashed line) particle spectrum in the silicon detector. Top (bottom) panel, runs at 400 (280) keV. The simulated spectrum is given in arbitrary units. Due to the strong preamplification, the spectra are affected by electronic noise below 1\,MeV.}
\end{figure}

Typical spectra taken with the silicon detector are shown in fig.\,\ref{fig:Silicon}, for the two $\alpha$-beam energies used here. The spectra show a broad peak at about 2.0\,MeV, corresponding to the most probable energy of the protons emitted by reaction R3, after $\approx$0.5\,MeV energy loss in the 25\,$\mu$m aluminum foil. The width of the peak is given by the kinematics and energy straggling in the Al foil. For $E_\alpha$ = 400\,keV, the high-energy cutoff is found at 3.0\,MeV, as expected from geometry and kinematics. 

It is evident that the energy of the proton peak due to reaction\,R3 is almost independent of beam energy, as it is dominated by the reaction $Q$-value. Similar considerations apply to the neutrons from reaction R2, which are not directly detected here. The Monte Carlo simulation reproduces the shape of the experimental data fairly well (fig.\,\ref{fig:Silicon}).

It is expected that some of the energetic deuterons created in the beam are implanted in the walls of the inner tube, at a depth of up to their 1.6\,$\mu$m range. There, they may again form a target for reactions R2 and R3. The protons created in nuclear reactions in this implantation layer will have slightly lower energy than protons created in the gas, due to the slowing of  the deuterons in this layer before the reaction takes place, and due to the proton energy loss (up to 0.07\,MeV) in the implantation layer. However, this effect is much smaller than the width of the peak.

During the experiment, it was found that the proton rate measured with the silicon detector increased with accumulated charge, by about 10\% for each step of 100\,C of charge. This confirms that the deuterium implanted in the target chamber actually does contribute to the proton yield, and by extension also the neutron yield. 

\section{Observations with the germanium $\gamma$-ray detector with and without beam}
\label{sec:GermaniumData}

\subsection{Adopted $\gamma$-ray regions of interest (ROIs)}

For the planned study of the $^2$H($\alpha$,$\gamma$)$^6$Li reaction with its very low expected yield, it is foreseen to use the highest beam energy available at LUNA, $E_\alpha$ = 400\,keV. For systematic checks, a second, lower beam energy is selected at 280\,keV, chosen so that the regions of interest (ROIs) do not overlap for the two reactions. 

The ROIs are determined as follows. The $^2$H($\alpha$,$\gamma$)$^6$Li reaction gives rise to a single $\gamma$ ray, that is observed in the detector at energy $E_\gamma$
\begin{equation} \label{gammarayenergy}
E_\gamma = Q + E \pm \Delta E_{\rm Doppler} - \Delta E_{\rm Recoil}
\end{equation}
The $\gamma$-ray energy shift due to the recoiling compound nucleus is negligible here, $\Delta E_{\rm Recoil}$ = 0.2\,keV. The Doppler correction, however, is significant, with the full Doppler shift amounting to $\Delta E_{\rm Doppler}$ $\approx$ 
16\,keV at $E_\alpha$ = 400\,keV. The energy loss over the length of the target amounts to about 2\,keV. 

As the target is extended over and beyond the full diameter of the germanium detector leading to emissions before and behind the detector (fig.\,\ref{fig:Setup}), the $\gamma$ rays, including their flanks, fall into a ROI that ranges from 1552.5-1581.5\,keV for $E_\alpha$ = 280\,keV, and from 1589.5-1624.3\,keV for $E_\alpha$ = 400\,keV. These ROIs will be assumed throughout the present work.

The respective contributions of electric dipole and electric quadrupole capture to the cross section are known only from theory \cite{Marcucci06-NPA,Hammache10-PRC}. Therefore, the angular distribution of the emitted $\gamma$ rays is highly uncertain. Because of the strong Doppler effect discussed above, the unknown angular distribution translates into an unknown shape of the $\gamma$ peak. In order to be insensitive to the transition type (dipole or quadrupole), the entire region of interest is used here.

\begin{figure*}[tb]
\includegraphics[angle=0,width=\textwidth,trim=6mm 0mm 6mm 0mm,clip]{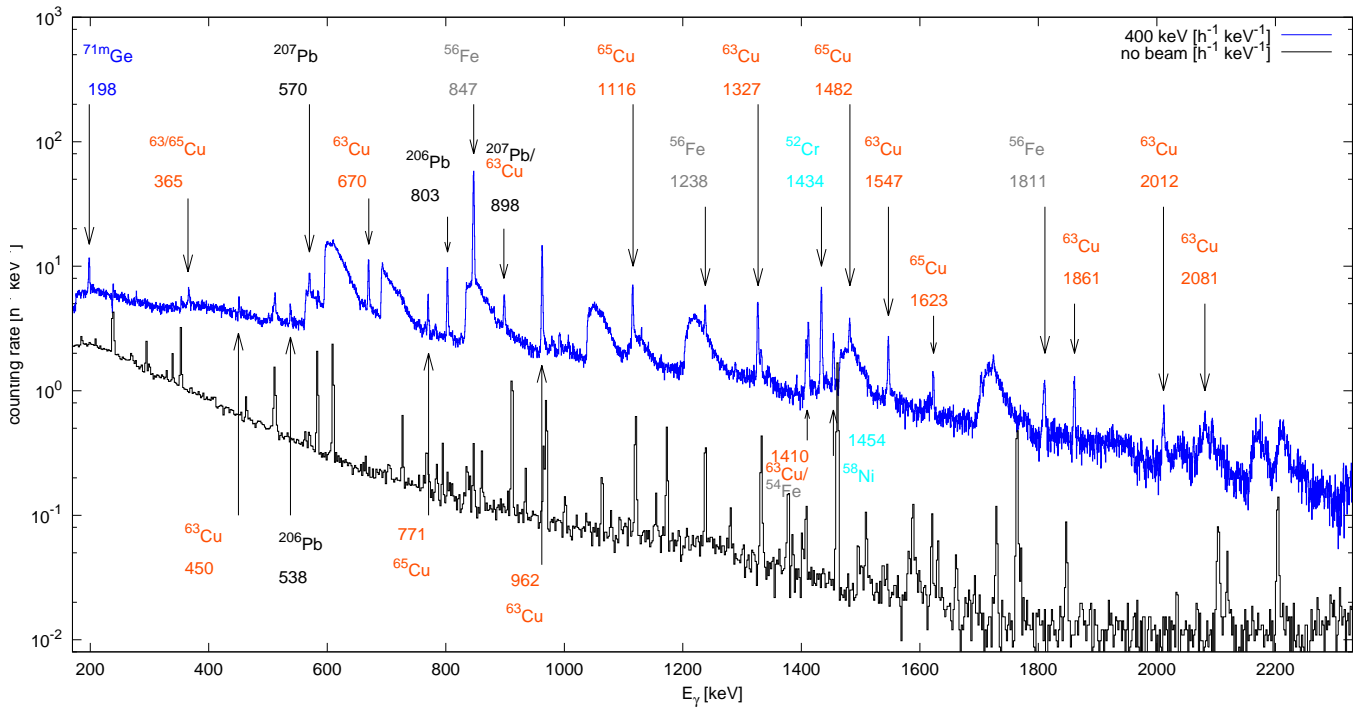}
\caption{\label{fig:400keV+labbg} Spectra taken with the germanium detector. Blue full line: in-beam spectrum at $E_\alpha$ = 400\,keV, $p_{\rm Target}$ = 0.3\,mbar,  laboratory background subtracted. The quantity plotted is the counting rate [h$^{-1}$ keV$^{-1}$]. In order to obtain the yield [C$^{-1}$ keV$^{-1}$], the values plotted have to be divided by 1.12\,C/h. Grey thin line: Laboratory background. The most important in-beam lines due to (n,n'$\gamma$) and (n,$\gamma$) processes on structural and shielding materials are marked with arrows, and the relevant target nuclide is given, as well as the $\gamma$-ray energy in keV. This information is also summarized in table~\ref{table:LevelsTransitions}. The asymmetric lines due to (n,n'$\gamma$) processes in the germanium detector itself are listed in table~\ref{table:NeutronInducedLines}.}
\end{figure*}

\begin{table}[bt]
 \centering
\begin{tabular}{|r|r|r@{~$\rightarrow$~}r|l|} \hline
\bf $E_\gamma$ [keV] & \multicolumn{1}{|l|}{\bf nuclide} & \multicolumn{2}{|l|}{\bf transition} & \bf Reference\\ \hline
198 & $^{71\rm m}$Ge & 198~$\rightarrow$~175 & 0 & \cite{Bunting74-NIM} \\ 
365 & $^{63}$Cu & 1327 & 962 & \cite{Dickens83-NPA}\\ 
366 & $^{65}$Cu & 1482 & 1115 & \cite{Dickens83-NPA}\\
450 & $^{63}$Cu & 1412 & 962 & \cite{Dickens83-NPA}\\ 
538 & $^{206}$Pb & 1341 & 803 & \cite{Dickens83-PRC} \\ 
570 & $^{207}$Pb & 570 & 0 & \cite{Kadi00-PRC} \\
670 & $^{63}$Cu & 670 & 0 & \cite{Dickens83-NPA}\\ 
771 & $^{65}$Cu & 771 & 0 & \cite{Dickens83-NPA}\\ 
803 & $^{206}$Pb & 803 & 0 & \cite{Dickens83-PRC} \\ 
847 & $^{56}$Fe & 847 & 0 & \cite{Lachkar74-NPA}\\
898 & $^{207}$Pb & 898 & 0 & \cite{Kadi00-PRC} \\
899 & $^{63}$Cu & 1861 & 962 & \cite{Dickens83-NPA}\\ 
962 & $^{63}$Cu & 962 & 0 & \cite{Dickens83-NPA}\\
1116 & $^{65}$Cu & 1116 & 0 & \cite{Dickens83-NPA}\\
1238 & $^{56}$Fe & 2085 & 847 & \cite{Lachkar74-NPA} \\
1327 & $^{63}$Cu & 1327 & 0 & \cite{Dickens83-NPA}\\
1408 & $^{54}$Fe & 1408 & 0 & \cite{Dickens73-NSE} \\
1412 & $^{63}$Cu & 1412 & 0 & \cite{Dickens83-NPA}  \\
1434 & $^{52}$Cr & 1434 & 0 & \cite{Karatzas78-NSE}\\
1454 & $^{58}$Ni & 1454 & 0 & \\
1482 & $^{65}$Cu & 1482 & 0 & \cite{Dickens83-NPA}\\
1547 & $^{63}$Cu & 1547 & 0 & \cite{Dickens83-NPA}\\
1623 & $^{65}$Cu & 1623 & 0 & \cite{Dickens83-NPA}\\
1811 & $^{56}$Fe & 2658 & 847 & \cite{Lachkar74-NPA}\\
1861 & $^{63}$Cu & 1861 & 0 & \cite{Dickens83-NPA}\\
2012 & $^{63}$Cu & 2012 & 0 & \cite{Dickens83-NPA}\\
2081 & $^{63}$Cu & 2081 & 0 & \cite{Dickens83-NPA} \\
 \hline
\end{tabular}
\caption{\label{table:LevelsTransitions} $\gamma$-ray energies, in keV, of observed Gaussian-shaped neutron-induced lines (arrows in fig.\,\ref{fig:400keV+labbg}), and the transition the lines originate from. See table\,\ref{table:NeutronInducedLines} for the broadened Ge(n,n'$\gamma$) peaks, and see the text for a discussion of the 198\,keV line.}
\end{table}

\subsection{Laboratory background}

Laboratory background induced by cosmic rays is so low inside the deep-underground LUNA facility that it is negligible for the purposes of the present study. However, this is not the case for background from environmental radionuclides, such as $^{40}$K and the nuclides of the $^{238}$U and $^{232}$Th decay chain. This latter background was attenuated by the lead shield (sec.\,\ref{subsec:Germanium}), and the laboratory background counting rate for 200\,keV $\leq$ $E_\gamma$ $\leq$ 2700\,keV was found to be reduced by a factor of 800 
compared to the unshielded setup. Inside the $^2$H($\alpha$,$\gamma$)$^6$Li region of interest, the reduction reaches a factor of 1100.

It has to be noted that a previous version of the present setup had achieved an even lower background level, mainly by lining the inside of the shield with 4\,cm of electrolytic copper, attenuating $\gamma$ rays and bremsstrahlung from the remaining $^{210}$Pb \cite{Caciolli09-EPJA}. For the present purposes, however, no copper liner was used, because of the high neutron capture cross section of copper. The lesser shielding with respect to Ref.\,\cite{Caciolli09-EPJA} leads to a factor of 16 higher laboratory background in the ROI, but it will be shown below that this no-beam background is still acceptable for the present purposes.

Two $\gamma$ rays in the remaining laboratory background fall inside the region of interest: The $^{228}$Ac line at 1588\,keV and the $^{212}$Bi line at 1621\,keV, both are part of the $^{232}$Th decay chain. The $^{232}$Th chain nuclides are believed to be an external contamination of the lead used for the shielding e.g. by dust particles.  

At higher $\gamma$-ray energies, the 2614 keV line of $^{208}$Tl is clearly visible: again a member of the $^{232}$Th decay chain. This line can in principle display a double escape peak inside the region of interest. However, the large size of the detector ensures that double escape peaks are negligible for all practical purposes. Even still, the Compton continuum caused by this line clearly influences the counting rate in the region of interest. The same is true for several lines from the $^{222}$Rn daughter $^{214}$Bi, in particular at 1730 and 1764\,keV. 

Due to a constant flushing of the anti-radon box with dry nitrogen, the laboratory background in the two \linebreak $^2$H($\alpha$,$\gamma$)$^6$Li ROIs was found to be stable within statistics over a period of several months, at a level of \linebreak (2.40$\pm$0.16)$\times$10$^{-2}$ and (3.32$\pm$0.17)$\times$10$^{-2}$ counts\,keV$^{-1}$\,h$^{-1}$ for the $E_\alpha$ = 280 and 400\,keV ROI, respectively.

In conclusion, the level of laboratory background is sufficiently low for the present purposes. It is stable in time and does not depend on the beam, so it can be subtracted.

\subsection{Structure of the background induced by the ion beam}

Due to the parasitic reactions R2 and R3, a certain amount of unwanted energetic neutrons, protons, tritons, and $^3$He particles are produced. They, in turn, may be captured by some structural, shielding or detector material. 

Energetic neutrons cause two main categories of effects in the setup. The first consists of capture and scattering effects on structural and shielding materials. They give rise to a number of well-defined $\gamma$ lines with Gaussian shape (fig.\,\ref{fig:400keV+labbg} and table\,\ref{table:LevelsTransitions}). The detector end-cap consists of electrolytic copper, and the beam stop also consists of copper. A number of lines from the two stable copper isotopes $^{63,65}$Cu are indeed observed (red markers in fig.\,\ref{fig:400keV+labbg}) \cite{Dickens83-NPA}. The gas target chamber, the collimator and the small tube inserted in the chamber consist of AISI 304 steel, and related neutron-induced lines from scattering on $^{54,56}$Fe \cite{Lachkar74-NPA}, $^{58}$Ni and $^{52}$Cr are apparent (grey and turquoise markers in fig.\,\ref{fig:400keV+labbg}). At 803\,keV, a line from neutron scattering on $^{206}$Pb from the massive lead shield surrounding the target chamber is observed.

\begin{table}[tb] 
 \centering
\begin{tabular}{|r|l|l|}
\hline
\multicolumn{1}{|l|}{$E_\gamma$ [keV]} & Reaction & Reference \\[0mm]
\hline
563 & $^{76}$Ge(n,n'$\gamma$) & \cite{Bunting74-NIM,Heusser93-NIMB,Ljungvall05-NIMA,Mei08-PRC} \\ 
596 & $^{74}$Ge(n,n'$\gamma$) & \cite{Bunting74-NIM,Heusser93-NIMB,Ljungvall05-NIMA,Mei08-PRC} \\ 
608 & $^{74}$Ge(n,n'$\gamma$) & \cite{Bunting74-NIM,Ljungvall05-NIMA,Mei08-PRC} \\ 
691 & $^{72}$Ge(n,n')$^{72}$Ge$^*$(IC) & \cite{Bunting74-NIM,Heusser93-NIMB,Ljungvall05-NIMA,Mei08-PRC} \\ 
834 & $^{70}$Ge(n,n'$\gamma$) & \cite{Bunting74-NIM,Heusser93-NIMB,Mei08-PRC} \\ 
1040 & $^{70}$Ge(n,n'$\gamma$) & \cite{Bunting74-NIM,Mei08-PRC} \\ 
1108 & $^{76}$Ge(n,n'$\gamma$) & \cite{Bunting74-NIM,Fehrenbacher96-NIMA} \\
1204 & $^{74}$Ge(n,n'$\gamma$) & \cite{Fehrenbacher96-NIMA} \\
1464 & $^{72,74}$Ge(n,n'$\gamma$) & \cite{Mei08-PRC} \\
1483 & $^{74}$Ge(n,n'$\gamma$) & \\
1697 & $^{74}$Ge(n,n'$\gamma$) & \\
1708 & $^{70}$Ge(n,n'$\gamma$) & \\
2155 & $^{70}$Ge(n,n'$\gamma$) & \\
2198 & $^{74}$Ge(n,n'$\gamma$) & \\
\hline
\end{tabular}
\caption{$\gamma$-ray energies, in keV, of  broadened Ge(n,n'$\gamma$) peaks in the 200-2200\,keV energy range (fig.\,\ref{fig:400keV+labbg}). The germanium nuclear level energies are taken from the ENSDF file \cite{ENSDF}. Previous observations of these lines are indicated, where they are known to the authors. The 691\,keV feature is caused by internal conversion (IC). See table\,\ref{table:LevelsTransitions} for the Gaussian-shaped neutron-induced lines.}
 \label{table:NeutronInducedLines}
\end{table}

\begin{figure*}[tb]
\includegraphics[angle=0,width=\textwidth]{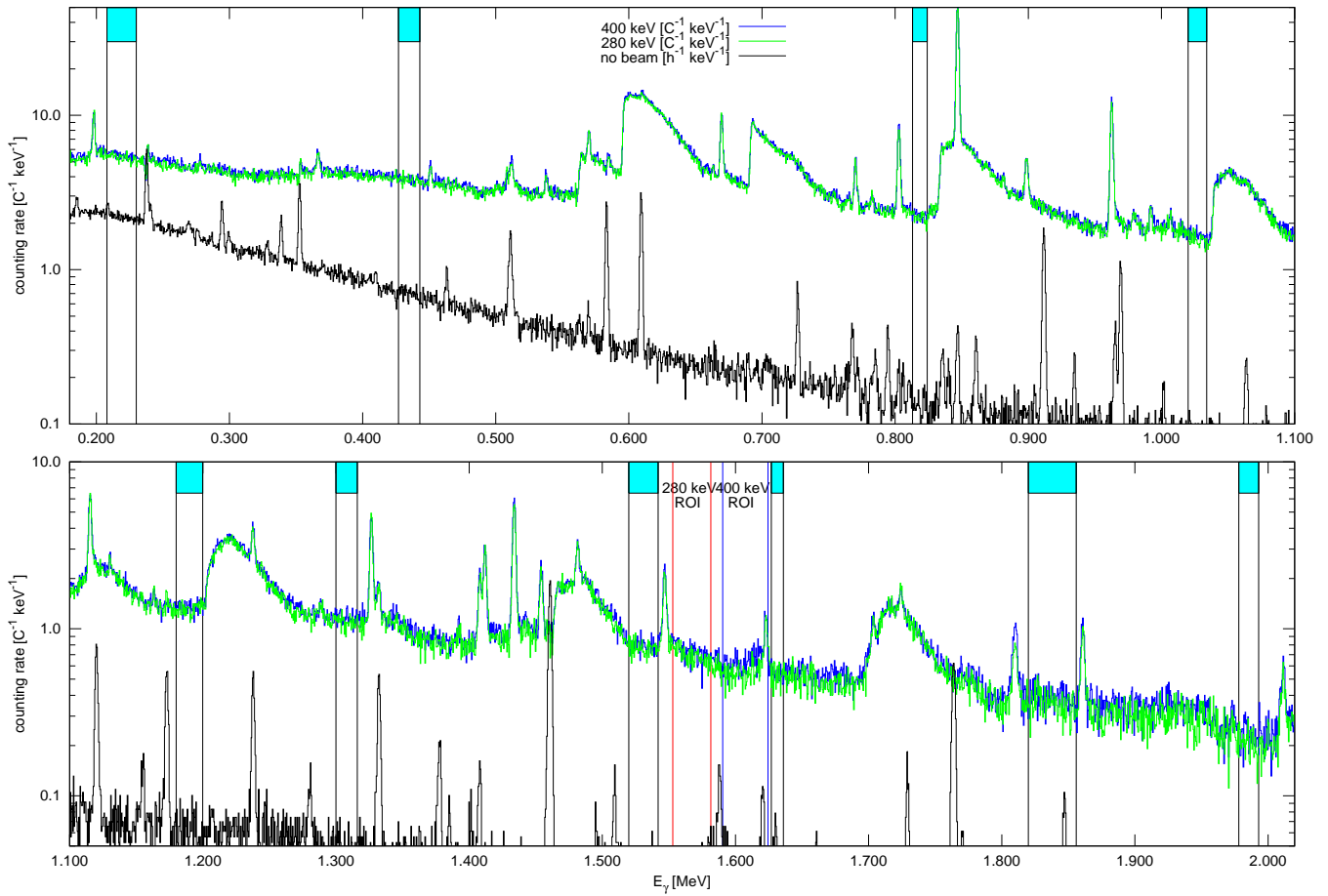}
\caption{\label{fig:Augenpulver} In-beam spectra at $E_\alpha$ = 400\,keV (blue line, laboratory background subtracted), $E_\alpha$ = 280\,keV (green line, laboratory background subtracted), and laboratory background (black line). The flat regions used for the empirical parameterization (sec.\,\ref{subsec:EmpiricalParameterization}, fig.\,\ref{fig:Yieldratio}) are marked by light blue boxes. The ROIs for $E_\alpha$ = 280\,keV (400\,keV) are indicated by red (blue) lines. For the in-beam data, the quantity plotted is the yield [C$^{-1}$ keV$^{-1}$]. In order to obtain the counting rate [h$^{-1}$ keV$^{-1}$], the values plotted have to be multiplied by 0.97\,C/h (0.89\,C/h) for $E_\alpha$ = 280 (400) keV. For the laboratory background, the counting rate [h$^{-1}$ keV$^{-1}$] is plotted.}
\end{figure*}

The second main effect is caused by (n,n'$\gamma$) processes of energetic neutrons on the germanium detector material itself. In addition to the emission of a $\gamma$ ray from the deexcitation of an excited nuclear state in the germanium target nucleus, the recoiling target nucleus also deposits energy in the detector material \cite{Ljungvall05-NIMA}, leading to a characteristic triangular shape starting at the energy of the  $\gamma$ ray. The broad features due to this effect (fig.\,\ref{fig:400keV+labbg}) can be attributed to a number of excited states in several stable germanium isotopes (table~\ref{table:NeutronInducedLines}). A number of the features observed here have been reported in previous work \cite{Bunting74-NIM,Heusser93-NIMB,Fehrenbacher96-NIMA,Ljungvall05-NIMA}

The $^{71\rm m}$Ge metastable state at $E_{\rm x}$ = 198\,keV is a special case. It is excited by the $^{70}$Ge(n,$\gamma$) reaction \cite{Bunting74-NIM,Heusser93-NIMB}. Due to its long half-life of 20\,ms, $^{71\rm m}$Ge decays only after the recoiling nucleus has been stopped.
The $^{71\rm m}$Ge level at 198\,keV decays via the 175\,keV state in $^{71}$Ge and onward to the ground state. The sum of these two $\gamma$ rays gives rise to a Gaussian peak at 198\,keV $\gamma$-ray energy in the spectrum (fig.\,\ref{fig:400keV+labbg}). 

In principle, further nuclear reactions in addition to the neutron-induced reactions discussed here are possible. 
In addition to energetic neutrons, reactions R2 and R3 give rise to 0.8-1.0\,MeV tritons and $^3$He particles. These particles, in turn, may in principle give rise to the \linebreak $^2$H(t,n)$^4$He and the $^2$H($^3$He,p)$^4$He reactions when passing the deuterium gas of the target, and to further triton- and $^3$He-induced reactions in the structural, shielding, and detector materials. Also (p,p'$\gamma$) reactions are in principle possible.
However, the cross sections for these charged particle induced reactions are lower than those for neutron scattering, and no clear signature has been found for them in the $\gamma$-spectra, so they are neglected here.


\subsection{Overall energy dependence of the ion-beam induced background}
\label{subsec:EmpiricalParameterization}

\begin{figure*}[tb]
\includegraphics[angle=0,width=\textwidth]{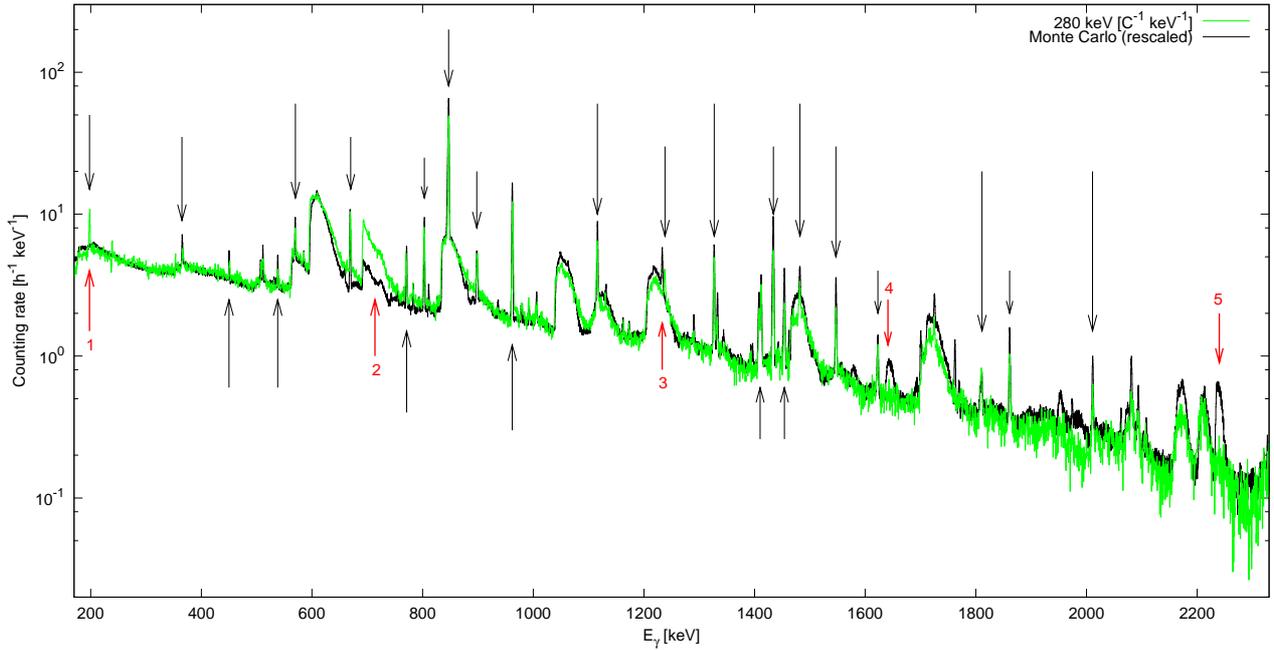} 
\caption{\label{fig:280keV+MonteCarlo} Comparison of the experimental spectrum at $E_\alpha$ = 280\,keV (green line) and the GEANT4 simulation for the same conditions (black line). The grey arrows are at the same positions as in fig.\,\ref{fig:400keV+labbg}, but the labels have been omitted here for clarity. The numbered red arrows denote the most important discrepancies between the measured and the simulated spectrum and are discussed in the text.}
\end{figure*}

The in-beam spectra display very similar features for the two beam energies $E_\alpha$ = 280 and 400\,keV. 
The reason for this is that the maximum neutron energy varies not very much with beam energy, from 3.3\,MeV at $E_\alpha$ = 400\,keV to 3.1\,MeV at $E_\alpha$ = 280 \,keV. 
The mutual similarity of the two neutron-induced spectra is actually even closer than the similarity between the experimental spectrum at \linebreak $E_\alpha$ = 280\,keV and the simulated spectrum at the same energy (fig.\,\ref{fig:280keV+MonteCarlo}). Based on this information, it seems useful to attempt an empirical parameterization of the energy dependence of the ratio between the counting rates at these two beam energies, after normalization for beam intensity. If successful, such a parameterization will in principle allow to use one of the two runs as a monitor run for the background of the other run, and vice versa, if the ROIs are disjunct.

The ion beam induced $\gamma$ rays discussed in the previous section all give rise to a Compton continuum. Near the $^2$H($\alpha$,$\gamma$)$^6$Li region of interest, this continuum is an order of magnitude higher than the laboratory background.

In the experimental spectrum at the two beam energies considered here, flat regions have been selected which have a width of at least 15\,keV and do not exhibit any significant structure (fig.\,\ref{fig:Augenpulver}). In these flat regions, the spectral shape is nearly identical for the two cases $E_\alpha$ = 280 and 400\,keV, and the ratio of the two respective yields is near unity. However, for higher beam energy also a slightly harder neutron spectrum is expected, so the yield ratio should exhibit an energy dependence. This is indeed the case (fig.\,\ref{fig:Yieldratio}). As expected, with increasing $\gamma$-ray energy the background yield at $E_\alpha$ = 400\,keV increases slightly with respect to the 280\,keV yield. 

This behavior has been parameterized empirically with a quadratic function (fig.\,\ref{fig:Yieldratio}). The quadratic function has been selected, because its $\chi^2/\nu$ value is 1.7, better than for a linear function which was also tried and which yields $\chi^2/\nu$ = 4.0. More complex fit functions have not been attempted here. 

\section{Discussion}
\label{sec:Discussion}

In the following text, the in-beam data and the simulation are compared, and the consequences for the planned $^2$H($\alpha$,$\gamma$)$^6$Li experiment at LUNA are discussed. 

\subsection{Comparison of in-beam data and simulation}
\label{subsec:Discussion_Inbeam}

The in-beam spectrum is plotted together with a Monte Carlo simulation for the $E_\alpha$ = 280\,keV case (fig.\,\ref{fig:280keV+MonteCarlo}). As the spectra at $E_\alpha$ = 280 and 400\,keV are very similar to each other (fig.\,\ref{fig:Augenpulver}), the following discussion also applies to the latter beam energy. 

The general features of the spectrum are rather well reproduced. Therefore, here only some notable but relatively minor discrepancies are discussed, following the numbering scheme in fig.\,\ref{fig:280keV+MonteCarlo}:
\begin{itemize}
\item[1.] The Gaussian sum peak denoting the $^{71\rm m}$Ge metastable state at $E_{\rm x}$ = 198\,keV \cite{Bunting74-NIM,Heusser93-NIMB} is observed in the experiment, but not in the simulation. This is due to the well known GEANT4 problem that while the decay of metastable states is correctly handled by the {\tt G4RDM} module, they are not correctly produced by the inelastic capture module. The 198\,keV state lives so long that the recoiling $^{71\rm m}$Ge is completely stopped long before it decays, so the typical neutron triangle structure does not apply. A peak at the same energy may also be caused by the $^{19}$F(n,n'$\gamma$) reaction, but this would require an unplausibly high amount of fluorine near the detector. 
\item[2.] The neutron triangle at 691\,keV is much smaller in the simulation than in the data. The 0$^+$ excited state at $E_{\rm x}$ = 691\,keV in $^{72}$Ge decays exclusively by internal conversion to the 0$^+$ ground state, meaning its energy is detected in the present large germanium detector with an efficiency close to 1. The Monte Carlo simulation erroneously treats this level as decaying by $\gamma$-ray emission, meaning the $\gamma$ ray is detected only with the intrinsic efficiency of the detector, which is considerably smaller than 1.
\item[3.] The 1238\,keV line assigned to the transition between the exited states at 2085 and 847\,keV in $^{56}$Fe shows up at 1233\,keV instead. It should be noted that the decay of the 847\,keV level to the ground state in $^{56}$Fe is reproduced at the correct energy. The observed 5\,keV $\gamma$-ray energy difference is too large to be explained by the Doppler effect, which would amount to less than 2\,keV in this case. This points to a possible error in the nuclear structure database included in GEANT4.
\item[4.] There is a feature at 1632\,keV in the simulation but not in the data. It looks like a Ge(n,n'$\gamma$) triangle but with smaller width, thus lower deposited germanium energy. It is therefore most probably due to the population of a relatively high-energy germanium level in the simulation but not in the data.

\item[5.] Same as 4. above, but at 2240\,keV.

\end{itemize}

Owing to the general good agreement between data and simulation, no significant discrepancies can be found in the Compton continuum, especially not near the \linebreak $^2$H($\alpha$,$\gamma$)$^6$Li ROIs. This confirms that the simplified model introduced in sec.\,\ref{sec:MonteCarlo} correctly describes the main processes in the experiment.

\begin{figure}[tb]
\includegraphics[angle=0,width=\columnwidth,trim=6mm 6mm 6mm 6mm,clip]{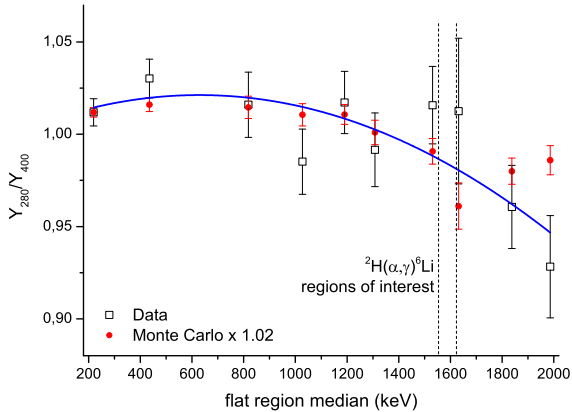} 
\caption{\label{fig:Yieldratio} Ratio of yields at $E_\alpha$ = 280 and 400\,keV, respectively. The data are from the flat regions marked in fig.\,\ref{fig:Augenpulver}. The empirical parameterization by a quadratic function is shown as a full line. The Monte-Carlo based ratios are scaled by a factor of 1.02 in order to match the high statistics data point at 200\,keV.}
\end{figure}

\subsection{Effects for the data analysis of a $^2$H($\alpha$,$\gamma$)$^6$Li experiment }
\label{subsec:Discussion_d+alpha}

If one assumes that the $^2$H($\alpha$,$\gamma$)$^6$Li cross section is equal to the recommended value from Ref.\,\cite{Hammache10-PRC}, at $E_\alpha$ = 400\,keV a signal to noise ratio of about 1:12 is obtained with the present neutron induced background. Therefore, it is important that the background level is understood with a precision on the percent level. The present analysis may lay the groundwork for such a precision. 

It has been shown above that the Monte Carlo simulation generally reproduces the experimentally observed spectrum rather well, both the general behavior and specific features (fig.\,\ref{fig:280keV+MonteCarlo}).  The ratio of the yields for the flat spectrum regions is in fair agreement with the data (fig.\,\ref{fig:Yieldratio}). 
The mutual agreement in shape is even better between the $E_\alpha$ = 280\,keV and 400\,keV runs, respectively. 

One remaining discrepancy between the $E_\alpha$ = 280 and 400\,keV runs lies in the counting rate for the 1811\,keV $^{56}$Fe line. It is higher at 400\,keV, possibly due to the somewhat higher neutron energy for this run. The Compton edge for this particular $\gamma$ ray lies exactly between the two ROIs for the $^2$H($\alpha$,$\gamma$)$^6$Li reaction. Therefore, the effects of the Compton continuum and multi-Compton events of the 1811\,keV $\gamma$ ray on these two ROIs has to be taken into account separately. Based on the data of the 1836\,keV $\gamma$ ray from the $^{88}$Y source, it is estimated that the Compton continuum of the 1811\,keV $\gamma$ ray leads to a correction of less than 1\% of the raw counts in each of the ROIs.

Similarly minor effects are expected if the counting rate at high energies, beyond the 0.2-2.0\,MeV energy range discussed in the present work, differs significantly between the two runs. Due to the limited statistics, this high $\gamma$-ray energy range is not considered further here.

The $^{65}$Cu line at 1623\,keV falls within the ROI for $E_\alpha$ = 400\,keV. It contributes about 6\% to the total counts in the ROI (fig.\,\ref{fig:Augenpulver}). This problem can be mitigated by excluding the region of this line from the ROI, and thus accepting a somewhat reduced statistics for the $\gamma$ ray to be detected. 

\section{Conclusions and summary}
\label{sec:Summary}

The effects of a weak flux of energetic neutrons -- induced by energetic deuterons from elastic scattering and the $^2$H(d,n)$^3$He reaction -- on an underground ultra-low level in-beam $\gamma$-ray spectroscopy setup has been studied experimentally at two different $\alpha$-beam energies, using data from a HPGe detector and a silicon particle detector. Monte Carlo simulations and tests with a weak Am-Be neutron source have been performed to assess the reliability of the results. 

Owing to the ultra-low background setting and the large size of the HPGe detector, a number of (n,n'$\gamma$) features in the background of lead-shielded HPGe detectors at $E_\gamma$ = 1.4-2.2\,MeV has been described experimentally for the first time. 

The reliability of a GEANT4-based simulation for neu\-tron-induced effects in a high-purity germanium detector has been studied, and remaining differences with the data have been discussed. As a result, the background in the planned $^2$H($\alpha$,$\gamma$)$^6$Li experiment is now well understood. 

The effects of the background given by the presently studied ultra-low neutron flux on the future $^2$H($\alpha$,$\gamma$)$^6$Li experiment have been evaluated, and a possible empirical parameterization of the in-beam background has been performed. It was shown that this parameterization holds the potential to use a run at a given beam energy as a background monitor for a run at a different beam energy, if the ROIs are disjunct. Even though the background is one order of magnitude larger than the expected $^2$H($\alpha$,$\gamma$)$^6$Li signal, the present data indicate that a positive measurement may be possible.

\section*{Acknowledgments}

The authors are indebted to Luca Gironi (INFN LNGS) for some initial Monte Carlo simulations, and to the mechanical and electronic workshops of LNGS for technical support. 
Financial support by INFN, FAI, DFG (BE 4100-2/1), NAVI (HGF VH-VI-417), and OTKA (K101328) is gratefully acknowledged.

\end{document}